\documentstyle[multicol,aps]{revtex}

\begin{document}
\draft
\title{
Universal relations in the finite-size correction terms 
of two-dimensional Ising models 
}

\author{Yutaka Okabe\cite{okabe} and Naoki Kawashima\cite{nao}}
\address{
Department of Physics, Tokyo Metropolitan University,
Hachioji, Tokyo 192-0397, Japan
}

\date{Received \today}

\maketitle

\begin{abstract}
Quite recently, Izmailian and Hu [Phys. Rev. Lett. 
{\bf 86}, 5160 (2001)] studied the finite-size correction terms 
for the free energy per spin and the inverse correlation length of 
the critical two-dimensional Ising model.  They obtained 
the universal amplitude ratio for the coefficients of two series. 
In this study we give a simple derivation of this universal 
relation; we do {\it not} use an explicit form of series expansion. 
Moreover, we show that the Izmailian and Hu's relation is reduced to 
a simple and exact relation between the free energy and the correlation 
length.  This equation holds at any temperature and has the same form 
as the finite-size scaling.
\end{abstract}

\pacs{PACS numbers: 75.10.-b, 05.50.+q}

\begin{multicols}{2}
\narrowtext

Universality and scaling are two basic concepts in 
the study of phase transitions and critical phenomena 
\cite{stanley71,cardy96}. 
The critical properties are universal to a large extent 
depending on a few parameters, such as the space 
dimensionality and the symmetry of the order parameter. 
Critical exponents, critical amplitude ratios, and 
scaling functions are examples of universal quantities 
\cite{brezin76,aharony91}. 
Finite-size scaling \cite{fisher70,cardy88} 
has been increasingly important, 
partly due to the progress in the theoretical understanding 
of finite-size effects, and partly due to the application 
to the analysis of simulational results.
Recently, more attention has been paid to the universality of 
finite-size scaling functions \cite{pf84} for both 
percolation models \cite{hlc95} and 
Ising models \cite{ok96}. 
In two dimensions the relevance of the finite-size properties 
to the conformal field theory is another source of 
interest \cite{bcna86}. 

Quite recently, Izmailian and Hu \cite{Izmailian01}
studied the finite-size correction terms for 
the free energy per spin and the inverse correlation length of 
critical two-dimensional (2D) Ising models 
\cite{onsager44,wannier50,husimi50}.  
They obtained the universal amplitude ratio 
for the coefficients of two series. 
Let us denote the free energy per spin and 
the inverse correlation length for $N \times \infty$ lattice 
as $f_N$ and $\xi_N^{-1}$, respectively. 
Then, Izmailian and Hu \cite{Izmailian01} obtained analytic expressions 
for the finite-size correction coefficients $a_k$ and $b_k$ 
defined by 
\begin{eqnarray}
 N (f_N-f_{\infty}) &=& \sum_{k=1}^{\infty} \frac{a_k}{N^{2k-1}}, 
 \label{a_k}
 \\
 \xi_N^{-1} &=& \sum_{k=1}^{\infty} \frac{b_k}{N^{2k-1}},
 \label{b_k}
\end{eqnarray}
at the criticality for the square (sq), honeycomb (hc), 
and plane-triangular (pt) lattices.  
Here, $f_{\infty}$ denotes the value of $f_N$ as $N \to \infty$. 
They found that
\begin{equation}
 \frac{b_k}{a_k} = \frac{2^{2k}-1}{2^{2k-1}-1}
 \label{ratio}
\end{equation}
for all of these lattices, that is, 
the amplitude ratio $b_k/a_k$ is universal.
They also obtained similar expansions for the critical
ground state energy $E^{(0)}$ and the first energy gap 
($E^{(1)}-E^{(0)}$) of the one-dimensional (1D) quantum XY model with 
uniform field \cite{katsura62} at the critical field,
and found that the amplitude ratios of two coefficients 
have the same values as Eq.~(\ref{ratio}).

The lowest-order correction terms $a_1$ and $b_1$ are 
related to the central charge $c$ and the magnetic 
scaling field $x_H$ in the conformal field theory 
by $a_1=c\pi/6$ and $b_1=2\pi x_H$, respectively.
The finite-size correction terms of 2D systems have also 
received current interest. 
Quieroz \cite{queiroz00} numerically studied the finite-size 
corrections to scaling of correlation lengths and free energies 
of the critical 2D Ising and 3-state Potts models. 
The finite-size corrections to the energy and specific heat 
of the critical 2D Ising model were also analyzed 
by Salas \cite{salas01}.  
Moreover, Salas and Sokal studied several universal amplitude ratios 
for the critical 2D Ising model \cite{salas00}. 

In this paper we will give a simple derivation of the universal 
amplitude ratio, Eq.~(\ref{ratio}). 
Instead of using an explicit form of expansion in $1/N$, 
which is available only at the criticality, we show 
a finite-size-scaling like relation which holds at an arbitrary 
temperature.  The universal amplitude ratio can be readily 
derived from this relation.

Let us start with comparing the free energy per spin and 
the inverse correlation length for size $N$ and $N/2$.
Assuming the expansion, Eq.~(\ref{a_k}), we have the expression 
for the difference of $f_N$ and $f_{N/2}$ as 
\begin{equation}
 f_{N/2}-f_{N} 
 = \sum_{k=1}^{\infty} \frac{a_k}{N^{2 k}}
 \Big\{ 2^{2k}-1 \Big\}.
 \label{diff_f}
\end{equation}
In a similar way, using Eq.~(\ref{b_k}), we have 
\begin{equation}
 \xi_{N/2}^{-1} - \xi_{N}^{-1} 
 = \sum_{k=1}^{\infty} \frac{b_k}{N^{2 k-1}} \Big\{ 2^{2 k-1} - 1 \Big\}.
 \label{diff_xi}
\end{equation}
From Eqs.~(\ref{diff_f}) and (\ref{diff_xi}), we get 
the following statement: 
If the relation 
\begin{equation}
 f_{N/2} - f_{N} = \frac{1}{N} (\xi_{N/2}^{-1} - \xi_{N}^{-1}) 
 \label{diff_two}
\end{equation}
is satisfied, we have the relation of two coefficients, 
Eq.~(\ref{ratio}).  If we can show Eq.~(\ref{diff_two}) directly, 
explicit expressions 
for the coefficients $a_k$ and $b_k$ are not necessary. 
Equation (\ref{diff_two}) is compatible to the finite-size scaling 
\cite{cardy96,cardy88} 
of the singular part of the free energy for $N \times \infty$ systems, 
$
 f_N \propto (1/N) \xi_N^{-1} \ (N \to \infty); 
$
namely, Eq.~(\ref{diff_two}) should hold asymptotically at the critical point.  
In fact, as we show in the following, Eq.~(\ref{diff_two}) holds 
{\it exactly} at {\it any} temperature for several models 
that belong to the 2D Ising universality class.

First, we deal with the 2D Ising model defined by the Hamiltonian
\begin{equation}
 \beta {\cal H} = -J\sum_{<ij>} s_i s_j,
 \label{Ising}
\end{equation}
where the Ising variable $s_i$ takes $\pm 1$, 
$\beta = (k_B T)^{-1}$, and the summation is taken over 
the nearest-neighbor pairs of sites for $N \times \infty$ lattices. 
Using a transfer matrix method \cite{kaufman59,domb60}, 
we can calculate the free energy per spin, $f_N$, and 
the inverse correlation length, $\xi_N^{-1}$, 
through the relations 
\begin{eqnarray}
 f_N &=& \frac{1}{\zeta N} \ln{\Lambda_0}, 
 \label{f}
 \\
 \xi_N^{-1} &=& \frac{1}{\zeta} \ln{(\Lambda_0/\Lambda_1)}, 
 \label{xi}
\end{eqnarray}
where $\Lambda_0$ and $\Lambda_1$ are the largest and 
the second largest eigenvalues of the transfer matrix.
Here, a geometric factor $\zeta$ is $1$, $2/\sqrt{3}$ and
$1/\sqrt{3}$ for sq, hc, and pt lattices, respectively.
Exact expressions for eigenvalues 
$\Lambda_0$ and $\Lambda_1$ are available for sq 
\cite{onsager44,kaufman59,domb60}, hc \cite{husimi50}, 
and pt \cite{wannier50} lattices.

Let us consider the Onsager solution for the sq lattice 
with periodic boundary conditions \cite{onsager44}.  
The two leading eigenvalues of the transfer matrix are given by 
\begin{eqnarray}
 \Lambda_0 &=& (2 \sinh 2J)^{N/2}
 \exp \Big( \frac{1}{2} \sum_{r=0}^{N-1} \gamma_{2r+1} \Big),
 \\
 \Lambda_1 &=& (2 \sinh 2J)^{N/2}
 \exp \Big( \frac{1}{2} \sum_{r=0}^{N-1} \gamma_{2r} \Big).
\end{eqnarray}
Here, $\gamma_r$ is implicitly given by
\begin{eqnarray}
 \cosh \gamma_r = \frac{\cosh^2 2J}{\sinh 2J} - \cos \frac{r\pi}{N}.
\end{eqnarray}
Then, we get the free energy per spin $f_N$, Eq.~(\ref{f}), 
and the inverse correlation length $\xi_N^{-1}$, Eq.~(\ref{xi}), as 
\begin{eqnarray}
 f_N &=& \frac{1}{2} \ln (2 \sinh 2J) + 
 \frac{1}{2N} \sum_{r=0}^{N-1} \gamma_{2r+1},
 \label{f_N}
 \\
 \xi_N^{-1} &=& 
 \frac{1}{2} \sum_{r=0}^{N-1} (\gamma_{2r+1}-\gamma_{2r}).
 \label{xi_N}
\end{eqnarray}
To express $f_{N/2}$ and $\xi_{N/2}$, we replace $N$ by $N/2$ 
and $\gamma_r$ by $\gamma_r'$ in Eqs.~(\ref{f_N}) and (\ref{xi_N}), 
respectively. 
Here $\gamma_r'$ is the value for size $N/2$, 
and is related to $\gamma_r$ through the relation 
\begin{equation}
 \cosh \gamma_r' = \frac{\cosh^2 2J}{\sinh 2J} - \cos \frac{r\pi}{N/2}
 = \cosh \gamma_{2r},
\end{equation}
that is, 
\begin{equation}
 \gamma_r' = \gamma_{2r}.
 \label{rel_gamma}
\end{equation}
Equation~(\ref{rel_gamma}) is the basic relation for our argument. 
Using the relation (\ref{rel_gamma}), we have 
\begin{eqnarray}
 f_{N/2} - f_{N} 
 &=& \frac{1}{2N} \Big\{ 2(\gamma_{2}+\gamma_{6}+ \cdots+\gamma_{2N-2})
 \nonumber \\
 &\quad& -(\gamma_{1}+\gamma_{3}+\cdots+\gamma_{2N-1}) \Big\}.
 \label{sq_f}
\end{eqnarray}
Similarly, we have
\begin{eqnarray}
 \xi_{N/2}^{-1}&-&\xi_{N}^{-1} \nonumber \\
 &=& 
 \frac{1}{2} \Big\{ (\gamma_{2}+\gamma_{6}+\cdots+\gamma_{2N-2}
             -\gamma_{0}-\gamma_{4}-\cdots-\gamma_{2N-4}) 
 \nonumber \\
 &\quad& - (\gamma_{1}+\gamma_{3}+\cdots+\gamma_{2N-1}
  -\gamma_{0}-\gamma_{2}-\cdots-\gamma_{2N-2}) \Big\} \nonumber \\
 &=&
 \frac{1}{2} \Big\{ 2(\gamma_{2}+\gamma_{6}+\cdots+\gamma_{2N-2})
 \nonumber \\
 &\quad& -(\gamma_{1}+\gamma_{3}+\cdots+\gamma_{2N-1}) \Big\}. 
 \label{sq_xi}
\end{eqnarray}
Comparing Eqs.~(\ref{sq_f}) and (\ref{sq_xi}), 
we arrive at the desired relation, Eq.~(\ref{diff_two}). 
This implies that the ratio for the finite-size correction coefficients 
$b_k/a_k$ is given by Eq.~(\ref{ratio}).

Next we consider the Ising model on the hc lattice.
The two leading eigenvalues of the transfer matrix 
are given by \cite{husimi50} 
\begin{eqnarray}
 \Lambda_0 &=&  (2 \sinh 2J)^N \exp \Big( \sum_{r=0}^{N/2-1} 
 \gamma_{2r+1} \Big),
 \label{hc_lambda0}
 \\
 \Lambda_1 &=&  (2 \sinh 2J)^N \exp \Big( 
 \sum_{r=1}^{N/2-1} \gamma_{2r} + \frac{\gamma_0+\gamma_{N}}{2} \Big). 
 \label{hc_lambda1}
\end{eqnarray}
In this case, $\gamma_r$ is implicitly given by
\begin{eqnarray}
 \cosh{\gamma_r} &=& \cosh 2J \cosh 2J^* - \sin^2 \frac{\pi r}{N}
 \nonumber\\
 &-& \cos \frac{\pi r}{N} \Big( \sinh^2 2J \sinh^2 2J^* - 
 \sin^2 \frac{\pi r}{N} \Big)^{1/2},
 \label{hc_gamma}
\end{eqnarray}
where $J^*$ is defined by $(\cosh 2J - 1)(\cosh 2J^* - 1) = 1$. 
For size $N/2$, we replace $N$ by $N/2$ and $\gamma_r$ 
by $\gamma_r'$ in Eqs.~(\ref{hc_lambda0}) and (\ref{hc_lambda1}). 
The relation of $\gamma_r'$, the value for size $N/2$, 
and $\gamma_r$ can be obtained by using Eq.~(\ref{hc_gamma}); 
we have the same relation as Eq.~(\ref{rel_gamma}), 
$
 \gamma_r' = \gamma_{2r}.
$
Starting from Eqs.~(\ref{f}) and (\ref{xi}) together with 
Eqs.~(\ref{hc_lambda0}) and (\ref{hc_lambda1}), with some algebra, 
we get 
\begin{eqnarray}
N \zeta(f_{N/2} - f_{N}) 
 &=& \zeta(\xi_{N/2}^{-1}-\xi_{N}^{-1}) \nonumber \\
 &=& 2(\gamma_{2}+\gamma_{6}+\cdots+\gamma_{N-2}) \nonumber \\
 &\quad&  -(\gamma_{1}+\gamma_{3}+\cdots+\gamma_{N-1}).
\end{eqnarray}
Thus, we again obtain the relation (\ref{diff_two}).
In this way, we have shown that the amplitude ratio $b_k/a_k$ 
is given by Eq.~(\ref{ratio}).
For the pt lattice, we may use the star-triangle transformation 
\cite{syozi72} to show Eq.~(\ref{diff_two}) 
from the result of the hc lattice without making an explicit calculation.

Some comments should be added here. 
We have shown the universality of the amplitude ratio, 
Eq.~(\ref{ratio}), without using the condition of the criticality. 
This means that the property of this universality holds not only 
at the critical point but also at any temperature in 2D Ising models. 
The ratio $b_k/a_k$ takes the universal value for all the temperatures. 
However, we should note that the expression for the inverse 
correlation length, Eq.~(\ref{xi}), is valid only for $T \ge T_c$. 
For $T<T_c$, $\Lambda_0$ and $\Lambda_1$ are degenerate in the 
thermodynamic limit, $N \rightarrow \infty$, which means 
the existence of the long-range order \cite{kaufman59,domb60}.
We should subtract the long-range order contribution from 
the correlation function when considering the correlation length. 
We may regard $b_k$ as the correction amplitude for the right-hand side 
of Eq.~(\ref{xi}) for $T<T_c$.

Izmailian and Hu \cite{Izmailian01} also studied another model 
which belongs to the 2D Ising universality class. 
The 1D quantum XY model with uniform field, 
whose Hamiltonian is given by 
\begin{eqnarray}
 {\cal H} = &-& \frac{1}{4} \sum_{n=1}^N 
 \Big[ (1+\gamma) \sigma^x_n \sigma^x_{n+1} + 
       (1-\gamma) \sigma^y_n \sigma^y_{n+1}  \Big]
 \nonumber \\
 &-& \frac{h}{2} \sum_{n=1}^N \sigma^z_n,
 \label{xy} 
\end{eqnarray}
was exactly solved by Katsura \cite{katsura62}. 
Here, $\sigma^x$, $\sigma^y$ and $\sigma^z$ are the Pauli matrices. 
For $0<\gamma \le 1$, there is a critical magnetic field 
$h_c =1$, and the phase transition of this model belongs to 
the 2D Ising universality class.  For $\gamma=1$ it is also called 
the 1D transverse Ising model. 
The Hamiltonian of Eq.~(\ref{xy}) is diagonalized by a Jordan-Wigner 
transformation as
\begin{equation}
 {\cal H} = - \sum_k \Lambda(k) \ 
 \Big( \eta_k^* \eta_k -\frac{1}{2} \Big)
\end{equation}
where $\eta_k^*$, $\eta_k$ are fermionic creation and annihilation 
operators and
\begin{equation}
 \Lambda(k) = \sqrt{ (\cos k + h)^2 + (\gamma \sin k)^2}. 
 \label{lambda}
\end{equation}
We should note that the choice of $k$ depends on the boundary 
condition. 
The ground state energy $E^{(0)}$ and the first energy gap 
$\Delta E = E^{(1)}-E^{(0)}$ of the quantum spin model, respectively, 
correspond to the free energy and inverse correlation length 
for the Ising model,
that is, 
$$
 N f_N \Longleftrightarrow - E_N^{(0)}
\quad {\rm and} \quad
 \xi_N^{-1} \Longleftrightarrow \Delta E_N.
$$
For the $N$ quantum spin systems with the periodic boundary 
condition, we have \cite{burkhardt85,henkel87}
\begin{eqnarray}
 E_N^{(0)} &=& - \frac{1}{2} \sum_{r=0}^{N-1} \gamma_{2r+1},
 \label{e0}
 \\
 \Delta E_N &=& - \frac{1}{2} \sum_{r=0}^{N-1} (\gamma_{2r} 
            - \gamma_{2r+1} ),
 \label{e_diff}
\end{eqnarray}
where we have used the notation 
$
 \gamma_r = \Lambda(r\pi/N). 
$
Let us consider the energy for size $N$ and 
that for $N/2$ as in the classical 2D Ising model. 
For size $N/2$, we replace $N$ by $N/2$ and $\gamma_r$ 
by $\gamma_r'$ in Eqs.~(\ref{e0}) and (\ref{e_diff}). 
Here, $\gamma_r'$ is again the value for $N/2$, and 
the relation of $\gamma_r'$ and $\gamma_r$ can be obtained 
by using Eq.~(\ref{lambda}). 
As a result, we have the same relation as Eq.~(\ref{rel_gamma}), 
that is,
\begin{equation}
 \gamma_r' = \gamma_{2r}.
 \label{rel_lambda}
\end{equation}
Then, we obtain the expression for the difference between 
$E^{(0)}_N$ and $E^{(0)}_{N/2}$ as
\begin{eqnarray}
 E^{(0)}_{N/2} - E^{(0)}_N 
 &=& - \frac{1}{2} 
   \Big\{ 2(\gamma_2+\gamma_6+\cdots+\gamma_{2N-2}) \nonumber \\
 &\quad&  -(\gamma_1+\gamma_3+\cdots+\gamma_{2N-1})
   \Big\}.
 \label{xy_e0}
\end{eqnarray}
For the first energy gap, 
we can also calculate the difference between the values 
for $N$ and $N/2$. Using Eq.~(\ref{rel_lambda}), 
with some algebra, we finally have
\begin{eqnarray}
 \Delta E_{N/2} - \Delta E_N 
 &=& \frac{1}{2} 
 \Big\{ 2(\gamma_2+\gamma_6+\cdots+\gamma_{2N-2}) \nonumber \\
 &\quad&  -(\gamma_1+\gamma_3+\cdots+\gamma_{2N-1})
 \Big\}.
 \label{xy_de}
\end{eqnarray}
From Eqs.~(\ref{xy_e0}) and (\ref{xy_de}), we have
\begin{equation}
 - ( E^{(0)}_{N/2} - E^{(0)}_N ) = 
 \Delta E_{N/2} - \Delta E_N, 
 \label{diff_quantum}
\end{equation}
which shows that the amplitude ratio of the finite-size correction 
coefficient for the ground state energy $(-E^{(0)})$, $a_k$, and 
that for the first energy gap $\Delta E$, $b_k$, is given 
by Eq.~(\ref{ratio}).

The amplitude ratio $b_k/a_k$ takes the same value as the classical 
case.  And this is not only at the critical field $h_c=1$ 
but also for all the magnetic fields.  However, it should be 
mentioned that for low field, $h<h_c$, $E_N^{(1)}$ does not 
necessarily give the first excited energy level.  We should 
interpret that $b_k$ is the correction amplitude for 
the right-hand side of Eq.~(\ref{e_diff}) for $h<h_c$.

To summarize, we have given a simple derivation of the 
universal amplitude ratio, Eq.~(\ref{ratio}).
We have shown the relation of Eq.~(\ref{diff_two}) and its quantum 
counterpart, Eq.~(\ref{diff_quantum}). 
Although these equations have the same form as the finite-size scaling
\cite{cardy96,cardy88}, they are exact and valid for an arbitrary temperature. 
One can perform an analytic calculation of 
each correction amplitude only at the criticality \cite{Izmailian01}.  
Our key relations are Eqs.~(\ref{rel_gamma}) and (\ref{rel_lambda}), 
and we have not used the condition of the critical point.

Izmailain and Hu \cite{Izmailian01} used a perturbed conformal 
field theory to understand the correction terms.  It is interesting 
to study directly the amplitude ratio by using the conformal field theory. 
The present study may give a hint for such a direction.

The finite-size properties depend on the boundary conditions.  
Recently, the effect of peculiar boundary conditions, such as 
the M\"obius strip and the Klein bottle, was studied for 
the 2D Ising model \cite{kaneda01,lu01}.  The effect of 
boundary conditions on the finite-size correction terms is 
an interesting subject to study, 
which is now in progress. 

In this paper, we have started from the expansion of the free energy
and the inverse correlation length in odd powers of $1/N$ 
in Eqs. (\ref{a_k}) and (\ref{b_k}).  Away from the critical point, 
correction terms in even powers of $1/N$ may appear.  Our argument is 
easily extended to such a case, and the main conclusion 
remains the same. 

We would like to thank N. Sh. Izmailian for informing us of their 
results prior to publication and for valuable discussions. 
Thanks are due to H. Takano and H. Otsuka for valuable discussions. 
This work was supported by a Grant-in-Aid for Scientific Research 
from the Ministry of Education, Science, Sports and Culture, Japan.

\end{multicols}

\end{document}